\documentclass[preprint,12pt]{elsarticle}
\usepackage[T1]{fontenc}
\usepackage[latin9]{inputenc}
\usepackage{geometry}
\usepackage[english]{babel}
\usepackage{listings}
\usepackage{fancyhdr}
\usepackage{ar}
\usepackage{cancel}
\usepackage{graphicx}
\usepackage[caption=false]{subfig}
\usepackage[fleqn]{amsmath}
\usepackage{mathtools}
\usepackage{amsfonts}
\usepackage{acronym}
\usepackage{amssymb}
\usepackage[export]{adjustbox}
\usepackage{psfrag}
\usepackage{color}
\usepackage{float}
\usepackage{subfloat}
\usepackage{hyperref}
\usepackage{enumerate}
\usepackage{pgfkeys}
\usepackage{smartdiagram}
\usepackage{rotating}
\usepackage{rotfloat}
\usepackage{tabulary}
\usepackage{natbib}
\usepackage{epstopdf}
\usepackage{stmaryrd}
\usepackage{setspace}
\usepackage{gensymb}
\usepackage{verbatim}
\usepackage{tikz}
\usetikzlibrary{arrows,shapes,calc,topaths}
\usetikzlibrary{decorations.pathreplacing}
\tikzstyle{line}=[draw]
\tikzstyle{arrow}=[draw, -latex]
\usetikzlibrary{intersections}
\usepackage{nomencl}

\makenomenclature
\newcommand\Rey{\mbox{\textit{Re}}}  % Reynolds number
\newcommand\Scm{\mbox{\textit{Sc}}}  % Reynolds number
\newcommand\Pran{\mbox{\textit{Pr}}} % Prandtl number, cf TeX's \Pr product

\begin{document}
\journal{International Journal of Heat and Fluid Flow}
\begin{frontmatter}

\title{Direct numerical simulation of developed \\ compressible flow in square ducts}
\author{Davide Modesti$^{1,2}$, Sergio Pirozzoli$^3$ and Francesco Grasso$^1$}

\address{$^1$Cnam-Laboratoire DynFluid, 151 Boulevard de L'Hopital, 75013 Paris}
\address{$^2$Department of Mechanical Engineering, The University of Melbourne, Victoria 3010,Australia}
\address{$^3$Dipartimento di Ingegneria Meccanica e Aerospaziale, Sapienza Universit\`a di Roma, via Eudossiana 18, 00184 Roma, Italia}
\begin{abstract}
We carry out direct numerical simulation of compressible square duct flow in the range of bulk Mach numbers $M_b=0.2-3$, and
up to friction Reynolds number $\Rey_\tau=500$. The effects of flow compressibility on the secondary motions are found to be negligible, 
as the typical Mach number associated with
the cross-stream flow is always less than $0.1$. As in the incompressible case, we find that
the wall law for the mean streamwise velocity applies with good approximation with respect to the nearest wall,
upon suitable compressibility transformation.
The same conclusion also applies to a passive scalar field, whereas the mean temperature
does not exhibit inertial layers because of nonuniformity of the aerodynamic heating.
We further find that the same temperature/velocity relation that holds for planar channels is applicable
with good approximation for square ducts, and develop a similar relation between temperature and 
passive scalars.
\end{abstract}
\end{frontmatter}

\section{Introduction}

Internal flows in square ducts are common in many engineering applications involving both incompressible and compressible flows.
Typical low-speed applications involve cooling, water draining and ventilation systems,
whereas at high speed the interest is mainly for air intakes
and wing/fuselage junctions. 
Square duct flow exhibits secondary motions in the cross-steam plane.
These were first experimentally observed
by~\citet{nikuradse_26,prandtl_27}, who invoked the occurrence of 
eight counter-rotating vortices bringing
high-momentum fluid from the duct core towards the corners
to explain the bending of the streamwise velocity isolines.
A considerable number of experiments and numerical 
simulations have been produced to explain 
the nature of secondary motions.
In particular the present authors have recently developed~\citep{pirozzoli_18,modesti_18},
a direct numerical simulation (DNS) dataset of square duct flow
in the friction Reynolds number range $\Rey_\tau^*=h/\delta_v^*\approx150-1000$
(where $h$ is the duct half side length, and $\delta_v^* = \nu/u_{\tau}^*$ is the viscous length scale
based on the mean friction velocity $u_{\tau}^* = \sqrt{\overline{\tau}_w}/\rho$),
the largest currently available in the literature.
Despite their effect in redistributing the wall
shear stress along the duct perimeter, we have shown that secondary motions do not have
large influence on the bulk flow properties, and the streamwise velocity field can
be characterized with good accuracy as resulting from the superposition of four flat
walls in isolation.
Furthermore, we showed that secondary motions contribute approximately $6\%$ of the total friction, 
and act as a self-regulating mechanism of turbulence whereby wall shear stress non-uniformities
induced by corners are equalized, and universality of the wall-normal velocity profiles
is established.
 
Regarding the compressible flow regime experimental and numerical studies are rather limited.
\citet{davis_86} investigated supersonic developing adiabatic flow at inlet Mach number $M=3.91$ and unit Reynolds 
number $\Rey/m = 1.8\times10^6$. 
They found that secondary motions develop as
in the incompressible case and that the transformed van Driest velocity profiles obey
the universal logarithmic wall law.
More recently, \citet{morajkar_16} carried out a series of experiments for supersonic duct flow at $M=2.75$,
$\Rey/\mathrm{m} = 8.9\times10^6$ using stereoscopic particle image velocimetry.
Similar to the incompressible case, they found that the velocity isolines bulge
towards the duct corners due to eight counter-rotating cross-stream vortices. 
The prediction of secondary motions is notoriously difficult for 
turbulence models, especially for those based on the eddy-viscosity hypothesis~\citep{bradshaw_87}.  
\citet{mani_13} carried out Reynolds averaged Navier-Stokes simulations 
of supersonic square duct flow
using different eddy viscosity models, showing that satisfactory prediction of secondary motions
is recovered using quadratic constitutive relations. 
\citet{vazquez_02} performed large-eddy simulation (LES) of compressible isothermal duct flow 
at bulk Mach number $M_b=u_b/c_w=0.5$ (with $u_b$ the bulk flow velocity and $c_w$ the speed of sound at the wall), 
both with cooled walls and only one heated wall. The cooled case showed good agreement with 
incompressible data available in the literature, indicating that compressibility effects are negligible at that 
Mach number, whereas higher intensity of the secondary motions was observed for the case with one heated wall.
\citet{vane_15} carried out wall-modeled LES corresponding to the experimental setup of~\citet{davis_86}, and found that
the development of the secondary eddies is strongly affected by the
wall shear stress distribution, and that they can significantly alter the primary, axial flow. 
 
Although the available studies of compressible duct flow seem to agree that
the structure of secondary motions is weakly affected by compressibility, 
the quantitative effect of Mach number variations on the flow is not fully understood yet. 
In particular, an important practical issue is the definition of the relevant effective Reynolds number
for comparison across Mach numbers, which is intrinsically related to the subject of compressibility transformations~\citep{morkovin_62}.
In plane channel flow, \citet{modesti_16} found that the compressibility transformation derived by
\citet[hereafter referred to as TL]{trettel_16} yields nearly perfect collapse of the wall-scaled velocity distributions 
in a wide range of Mach numbers.
Understanding the behavior of passive scalars in compressible flow is important 
in particular to understand mixing processes in turbulent combustion.
However, passive scalars in compressible wall-bounded flows have received little attention so far, mainly limited 
to the case of planar channels~\citep{foysi_05} and pipe flow~\citep{ghosh_08,ghosh_10}.
Another important topic in compressible flows is represented by temperature/velocity relations~\citep{smits_06}.
Whereas these relations are well established in canonical flows~\citep{zhang_14}, their validity has never been verified for more complex geometries.
The aim of the present work is three-fold.
First, we attempt to extend the compressibility velocity transformations developed for plane channel flow to the case of multiple walls,
through suitable definition of the relevant effective Reynolds number.
Second, we derive a compressibility transformation for passive scalars.
Third, we analyze the temperature field with the main objective of verifying the 
validity of temperature/velocity relations.
Hence, we perform DNS of isothermal square duct flow in the range of bulk Mach number $M_b=0.2-3$, up to $\Rey_\tau^*=500$.

\section{Computational setup}

\begin{table*}
\begin{center}
\scalebox{0.8}{
\begin{tabular}{lccccccccccccccc}
\hline
Case &  $M_b$ & $\Rey_b$ & $\Rey_{\tau}^*$ & ${\Rey_{\tau}}_T^*$ &$N_x$ & $N_y$ & $N_z$ &$\Delta x^*$ & $\Delta z^*$ & $\Delta y_w^*$& $M_{\tau}$&$T_{\tau}^*/T_w$&${\Delta t}_{av}^*u_\tau^*/h$\\
\hline
 D02  &$0.2$  &$4410$ & $152$  & $146$ & $512$  & $128$ & $128$ & $5.6$&$3$ &$0.68$ & $0.014$ &$0.001$& $2290$\\
 D15A    &$1.5$  &$6000$  & $228$  & $141$ & $512$  & $128$ & $128$  &$8.4$& $5.4$ & $0.58$  & $0.082$ & $0.05$ &$830$\\
 D15B    &$1.5$  &$14600$ & $507$  & $332$ & $1024$  & $256$ & $256$ &$9.3$ &$6.0$ &$0.76$ & $0.075$ & $0.045$ & $1036$\\
 D3      &$3  $  &$9760$  & $483$  & $145$ & $1024$ & $256$ & $256$  &$8.9$ &$5.8$ &$0.61$ & $0.12$ & $0.14$ & $213$\\
\hline
\end{tabular}
}
\end{center}
\caption{Compressible duct flow dataset.  $M_b=u_b/c_w$ and $\Rey_b=2\rho_w u_b h/\mu_w$, are the bulk Mach and Reynolds number respectively;
        $\Rey_{\tau}^*=h/\delta_v^*$ and ${\Rey_{\tau}}_T^* = y_T^*(h)/\delta_v^*$ are the standard
        and transformed friction Reynolds number, as defined in Eqn.~\eqref{eq:TL}. 
        $N_i$ is the number of mesh points in the $i-$th
        direction, $M_\tau=u_\tau^*/c_w$ is the friction Mach number, $T_\tau^*$ is the global friction temperature,
        defined in equation~\eqref{eq:ttau}.
%       $M_{\max}^{yz}=\max((\overline{v}+\overline{w})/\overline{c})$ is the maximum Mach number in
%       the cross stream plane.
        $\Delta x$ is the mesh spacing in the streamwise direction, and
$\Delta z$, $\Delta y_w$ are the maximum and minimum mesh spacings
in the cross-stream direction, all given in global wall units, $\delta_v^*=\nu_w/u_{\tau}^*$.
        The box dimensions are $6\pi h\times 2h\times 2h$
        for all the flow cases. ${\Delta t}_{av}^*$ is the effective time averaging interval.}
\label{tab:testcases_duct}
\end{table*}

We solve the compressible Navier-Stokes equations for a perfect shock-free heat-conducting gas augmented with the transport equation for a passive scalar $\phi$,
\begin{subequations}
 \begin{align}
  \frac{\partial \rho}{\partial t}   + 
  \frac{\partial\rho u_j}{\partial x_j}  & = 0 ,
  \label{eq:mass}\\
  \frac{\partial \rho u_i}{\partial t}  +
  \frac{\partial\rho u_i u_j}{\partial x_j} &= -
  \frac{\partial p}{\partial x_i}           +
  \frac{\partial\sigma_{ij}}{\partial x_j}  +
  \Pi \delta_{i1} ,
  \label{eq:momentum}\\
  \frac{\partial \rho s}{\partial t} +
  \frac{\partial\rho u_j s}{\partial x_j} &= 
   \frac{1}{T}\left(-\frac{\partial q_j}{\partial x_j}       +
   \sigma_{ij}\frac{\partial u_i}{\partial x_j}\right)
  \label{eq:entropy},\\
  \frac{\partial \rho \phi}{\partial t} +
  \frac{\partial\rho u_j \phi}{\partial x_j} &= 
   \frac{\partial}{\partial x_j} \left( \rho\alpha \frac{\partial \phi}{\partial x_j}\right) + \Phi, 
  \label{eq:scalar}
 \end{align}
\end{subequations}
where $u_i$ is the velocity component in the i-th direction, $\rho$ is the fluid density,
$p$ is the pressure, $s=c_v\ln{(p\rho^{-\gamma})}$ is the entropy per unit mass,
$\gamma=c_p/c_v=7/5$ is the specific heat ratio,
$\sigma_{ij}$ and $q_j$ are, respectively, the 
viscous stress and the heat flux components,
\begin{equation}
 \sigma_{ij}=\mu\left(\frac{\partial u_i}{\partial x_j} + \frac{\partial u_j}{\partial x_i} -\frac{2}{3} \frac{\partial u_k}{\partial x_k} \delta_{ij}\right), 
\end{equation}
\begin{equation}
 q_j=-k\frac{\partial T}{\partial x_j}. 
\label{eq:heat_vec}
\end{equation}
The dependence of the viscosity coefficient on temperature is accounted for through Sutherland's law,
and the thermal conductivity is defined as $k=c_p\mu/\Pran$, with $\Pran=0.71$.
The forcing term $\Pi$ in equation~\eqref{eq:momentum}
is evaluated at each time step in order to discretely enforce constant mass-flow-rate in time, hence
the bulk Mach number is also constant. 
The passive scalar diffusivity
is $\alpha= \mu / \rho \Scm$, with $\Scm$ the Schmidt number, and the forcing term 
$\Phi$ in equation~\eqref{eq:scalar} is evaluated at each time step
to keep a constant scalar flow rate in time.
The equations are numerically solved using a fourth-order co-located
finite-difference solver, and the convective terms are discretized in such a way that
the total kinetic energy is preserved from convection in the inviscid limit~\citep{pirozzoli_10}. 
Viscous terms are expanded to Laplacian form and discretized using standard central finite-difference approximations.
The use of the entropy equation~\eqref{eq:entropy} in place of the total energy equation
is instrumental to semi-implicit time advancement, 
thus discarding the severe acoustic time step limitation in the wall-normal direction~\citep{modesti_18}. 
The equations are solved in a box 
of size $6\pi h\times2h\times 2h$, 
which was found to yield satisfactory insensitivity of the flow statistics~\citep{pirozzoli_18}.
Periodicity is enforced in the
streamwise direction, whereas isothermal no-slip boundary conditions are used at the walls
and implemented as described by~\citet{modesti_16}. Homogeneous Dirichlet
boundary conditions are used for the passive scalar variable.
The velocity field is initialized with the incompressible laminar solution
with superposed synthetic perturbations obtained through the digital filtering technique~\citep{klein_03}.
Density and temperature are initially uniform, whereas the passive scalar
is initialized as the streamwise velocity field, upon suitable rescaling.
Streamwise averaged statistics have been collected
at equal time intervals, and convergence of the flow statistics 
has been checked a-posteriori.
As observed in previous DNS studies of duct flow~\citep{gavrilakis_92,pirozzoli_18},
the time integration intervals needed to achieve statistical convergence
are much longer than those typical of plane channel flow, see table~\ref{tab:testcases_duct}. 
Supersonic simulations have been carried out at $M_b=1.5$ and $M_b=3$,
at $\Rey_\tau^*=220-500$ and $\Rey_\tau^*=500$, 
respectively (see table~\ref{tab:testcases_duct}).
A reference low-speed simulation at $M_b=0.2$, $\Rey_\tau^*=150$
has also been carried out (case D02 of table~\ref{tab:testcases_duct}),
which was shown to yield excellent agreement of mean and r.m.s. velocity 
with reference incompressible DNS data~\citep{pirozzoli_18}.
 
For the forthcoming analysis, the results are reported both in local and global wall units.
Accordingly, we introduce reference friction values for velocity, temperature, and passive scalar, namely
\begin{eqnarray}
u_{\tau}^2 = \nu_w \left. \frac{\partial \widetilde{u}}{\partial y}\right\rvert_w , \quad
\quad 
{u_{\tau}^*}^2 = \frac {h \overline{\Pi}}{2 \rho_w^*}, \label{eq:utau} \\
T_\tau = \frac{k_w}{\rho_w c_p u_\tau} \left.\frac{\partial \widetilde{T}}{\partial y}\right\rvert_w, \quad
T_\tau^*= \frac{h \overline{\Psi}}{2 \rho_w^* c_p u_\tau^*}, \label{eq:ttau} \\
\phi_{\tau}=\frac{\alpha_w}{u_\tau}\left.\frac{\partial \widetilde{\phi}}{\partial y}\right\rvert_w, \quad
\phi_\tau^* = \frac{h \overline{\Phi}}{2 \rho_w^*u_\tau^*}, \label{eq:phitau} \\
\end{eqnarray}
where quantities denoted with the $*$ superscript are averaged over the duct perimeter,
and $\Psi$ is the viscous dissipation function, to be defined in equation~\eqref{eq:temp_duct}.
For clarity of notation, hereafter $x$ denotes the streamwise direction, and
$y$ and $z$ the cross-stream and wall-normal directions,
and $u$, $v$ and $w$ are the respective velocity components.
Both Reynolds ($\phi=\overline{\phi} + \phi'$) and Favre ($\phi=\widetilde{\phi} + \phi''$,
$\widetilde{\phi}=\overline{\rho\phi}/\overline{\rho}$) decompositions will be considered in the following,
where the overline symbol denotes averaging in the streamwise directions and in time.
Accordingly, the Reynolds stress components are denoted as $\tau_{ij} = \overline{\rho}\widetilde{u''_i u''_j}$. 
 
\section{Results} \label{sec:results}

\subsection{Velocity field} \label{sec:transformations}

\begin{figure}
 \begin{center}
  \includegraphics[]{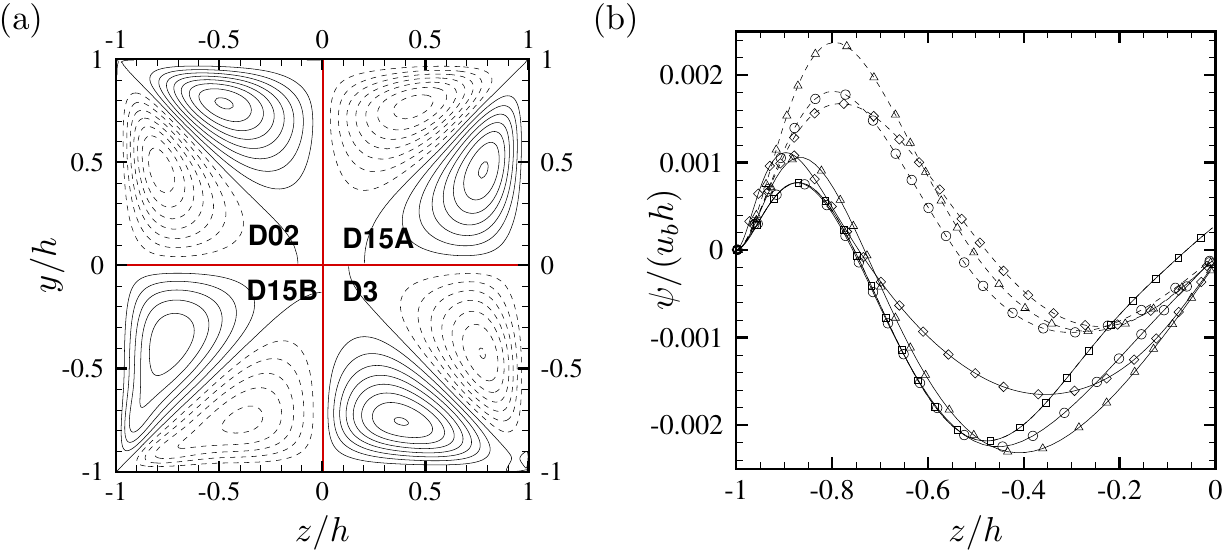}
  \vskip 1em
  \caption{Panel (a), contours of the streamfunction $\psi$ in the range $-0.002\le\psi/(u_bh)\le0.002$,
           in intervals of $0.0025$ (dashed lines denote negative values). Data are reported for flow cases
           D02 (top left), D15A (top right), D15B (bottom left), D3 (bottom right).
           Panel (b), 1D profiles at $y/h=-0.75$ (solid) and $y/h=-0.5$ (dashed)
  for case D02 (deltas), D15A (circles), D15B (diamond), D3 (squares).
           }
  \label{fig:psi_2d}
 \end{center}
\end{figure}

\begin{figure}
 \begin{center}
  \includegraphics[]{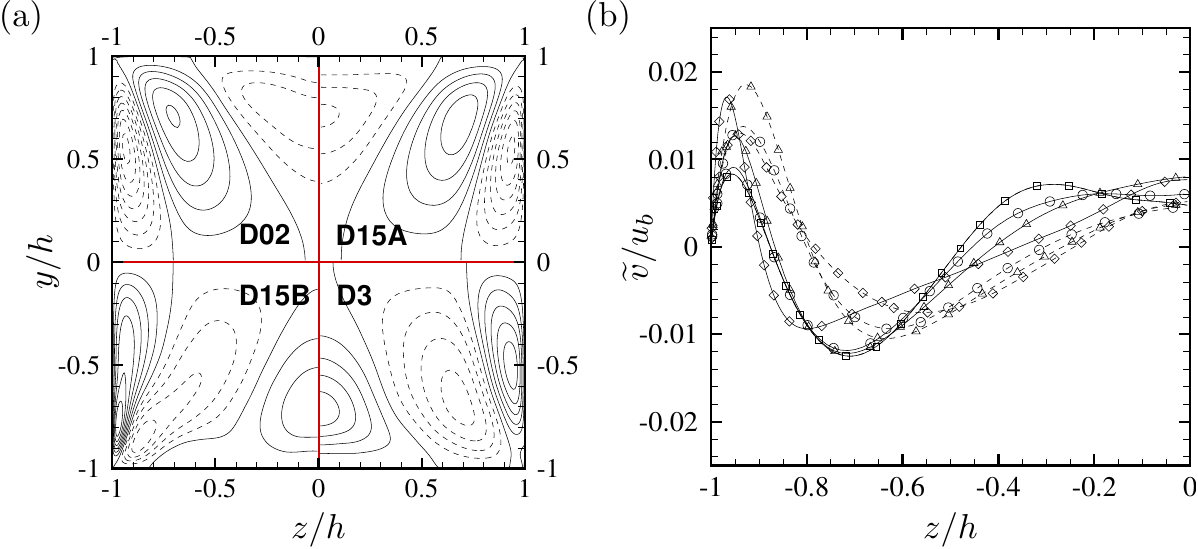}
  \vskip 1em
  \caption{Panel (a), contours of the mean cross-stream velocity component $\widetilde{v}$ in the range $-0.02\le\widetilde{v}/u_b\le0.02$,
           in intervals of $0.0025$ (dashed lines denote negative values). Data are reported for flow cases
           D02 (top left), D15A (top right), D15B (bottom left), D3 (bottom right). 
           Panel (b), 1D profiles at $y/h=-0.75$ (solid) and $y/h=-0.5$ (dashed)
  for case D02 (deltas), D15A (circles), D15B (diamond), D3 (squares).
           }
  \label{fig:mean_vv_2d}
 \end{center}
\end{figure}
 
In this section we analyze the structure of the mean velocity field including the secondary motions, with special 
reference to establishing the effect of compressibility on the validity of compressibility transformations for the wall law.
The structure of the secondary motions is hereafter analyzed by introducing a cross-flow stream function $\psi$, defined such
that at any point over the duct cross section
\begin{equation}
 -\overline{\rho} \widetilde{v}= \overline{\rho}_w\frac{\partial {\psi}}{\partial z},\quad
  \overline{\rho} \widetilde{w}= \overline{\rho}_w\frac{\partial {\psi}}{\partial y},
 \label{eq:psi}
\end{equation}
which satisfies mass conservation in the cross-stream plane.
In figure~\ref{fig:psi_2d} we show $\psi$ in a quarter of the duct, 
scaled with respect to $u_b$ and $h$, for the various flow cases of table~\ref{tab:testcases_duct}.
All the flow cases exhibit the same typical flow topology with eight counter-rotating eddies, 
which act to feed the low-momentum regions created at the corners. 
Figure~\ref{fig:mean_vv_2d} further shows that the cross-stream velocity component 
is characterized by a three-lobe structure, as in the incompressible case~\citep{pirozzoli_18}. 
The cross-stream velocity peaks are found to scale reasonably well with the bulk flow velocity, 
regardless of Mach and Reynolds number, with maximum value of about $2\%$ of $u_b$, which
is similar to the incompressible case~\citep{pirozzoli_18}.

\begin{figure}
 \begin{center}
  \includegraphics[]{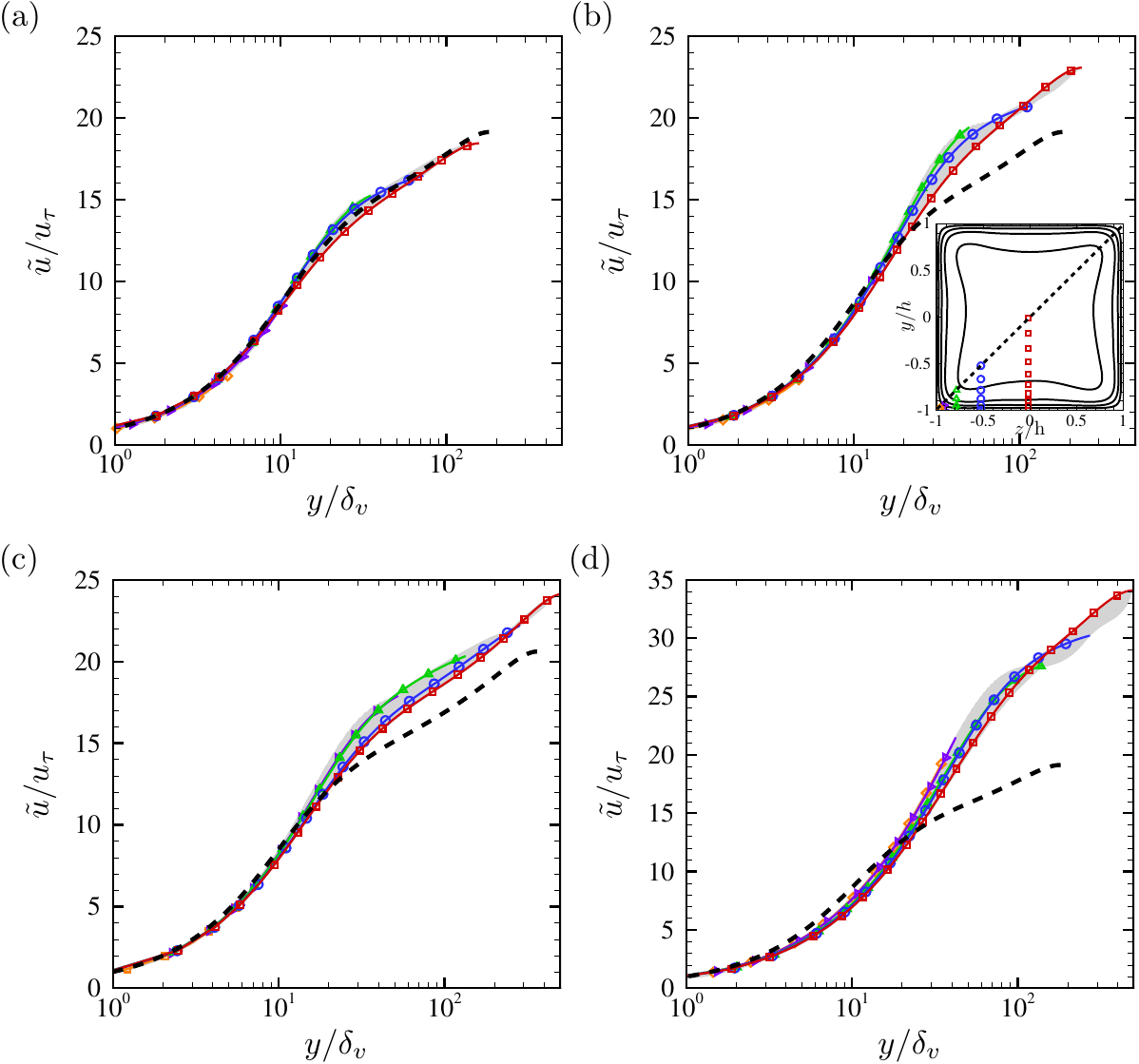}
  \vskip 1em
  \caption{Mean streamwise velocity profiles along the $y$ direction (up to the corner
bisector), given in local wall units at all $z$, 
for flow case D02 (a), D15A (b) D15B (c) and D3 (d).
Representative stations along the bottom wall are highlighted, namely $z^* = 15$ (diamonds), $(z+h)/h = 0.1$ (right triangles), $(z+h)/h = 0.25$
(triangles), $(z+h)/h = 0.5$ (circles), $(z+h)/h = 1$ (squares). The dashed lines denote mean profiles
from DNS of pipe flow at $\Rey_\tau = 180$ (a-b-d) and $\Rey_\tau=360$ (c) from~\citep{khoury_13}, 
The inset panel (b) shows the mean streamwise velocity in the cross-stream plane with
symbols denoting representative sections.
           }
  \label{fig:uu_untr}
 \end{center}
\end{figure}
\begin{figure}
 \begin{center}
  \includegraphics[]{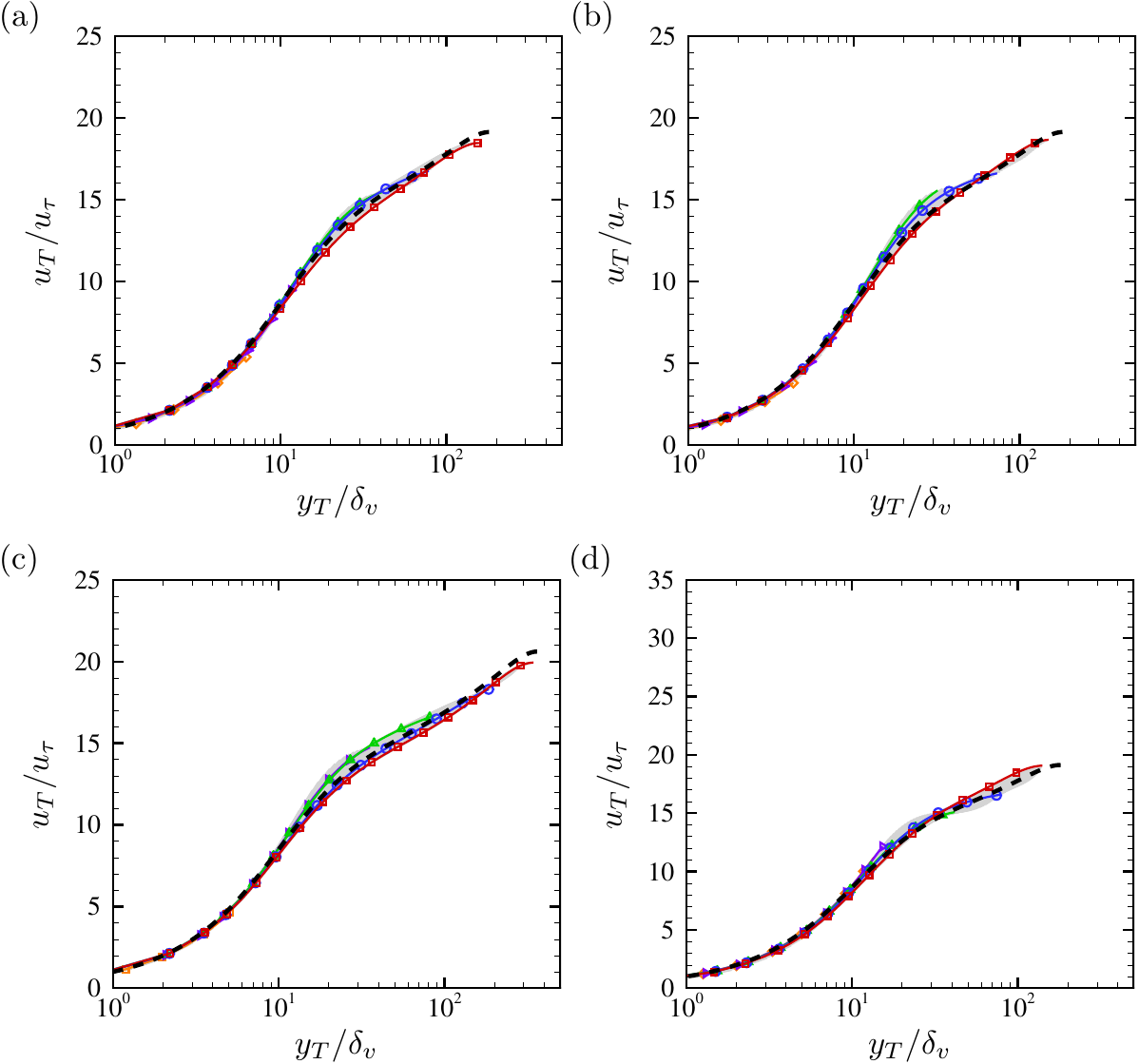}
  \vskip 1em
  \caption{Mean streamwise velocity profiles transformed according to equation~\eqref{eq:TL}
           along the $y$ direction (up to the corner
bisector), given in local wall units at all $z$,
for flow case D02 (${\Rey_{\tau}}_T^*=146$, panel a), D15A (${\Rey_{\tau}}_T^*=141$, panel b) 
D15B (${\Rey_{\tau}}_T^*=332$, panel c) and D3 (${\Rey_{\tau}}_T^*=145$, panel d).
Representative stations along the bottom wall
are highlighted, namely $z^* = 15$ (diamonds), $(z+h)/h = 0.1$ (right triangles), $(z+h)/h = 0.25$
(triangles), $(z+h)/h = 0.5$ (circles), $(z+h)/h = 1$ (squares). 
The dashed lines denote mean profiles
from DNS of pipe flow at $\Rey_\tau = 180$ (a,b,d) and $\Rey_\tau=360$ (c)~\citep{khoury_13}. 
           }
  \label{fig:uu_trett}
 \end{center}
\end{figure}

\begin{figure}
 \begin{center}
  \includegraphics[]{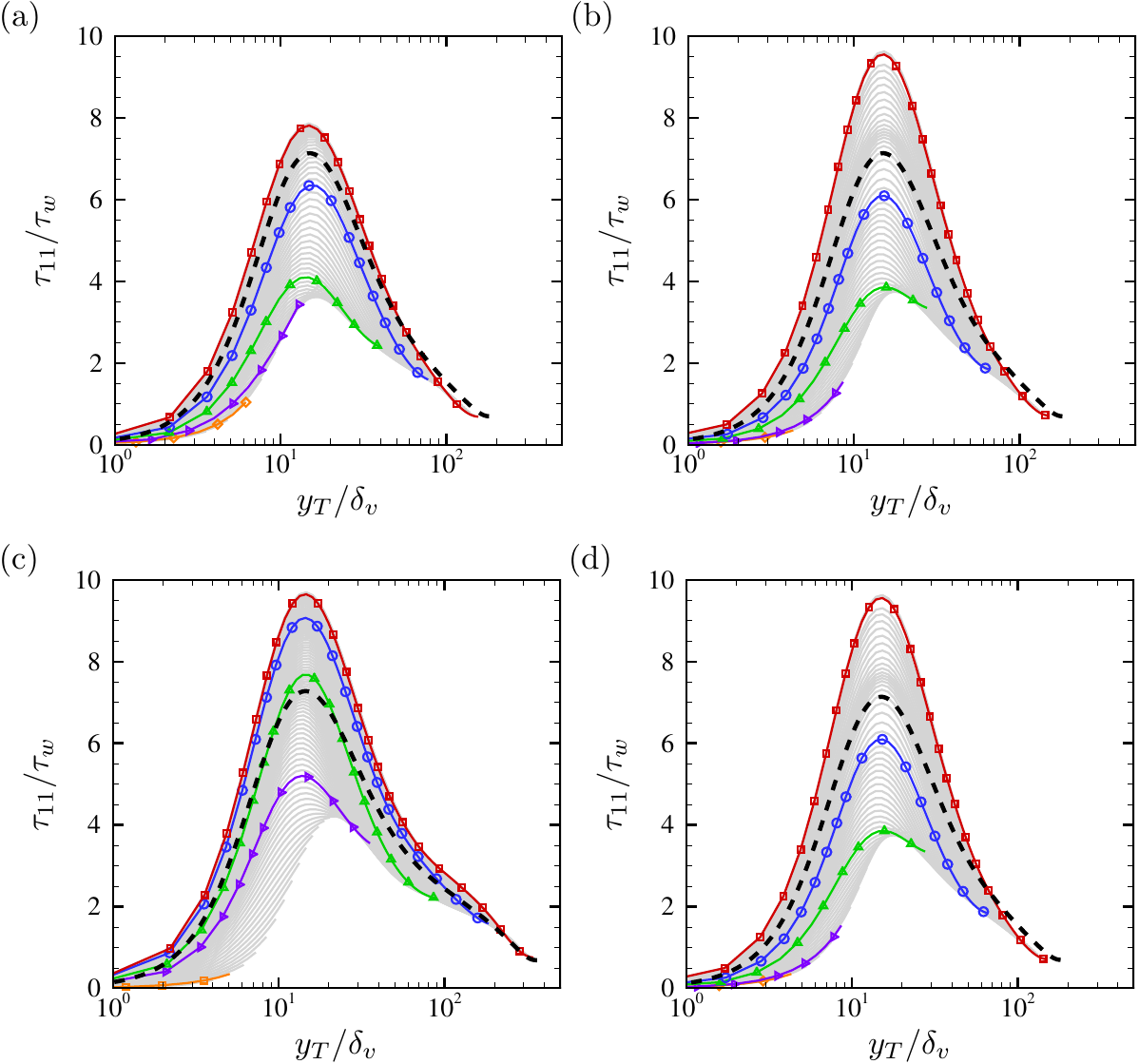}
  \vskip 1em
  \caption{Streamwise turbulent stresses
           along the $y$ direction (up to the corner
bisector), given in local wall units at all $z$,
for flow case D02 (${\Rey_{\tau}}_T^*=146$, panel a), D15A (${\Rey_{\tau}}_T^*=141$, panel b) 
D15B (${\Rey_{\tau}}_T^*=332$, panel c) and D3 (${\Rey_{\tau}}_T^*=145$, panel d).
Representative stations along the bottom wall
are highlighted, namely $z^* = 15$ (diamonds), $(z+h)/h = 0.1$ (right triangles), $(z+h)/h = 0.25$
(triangles), $(z+h)/h = 0.5$ (circles), $(z+h)/h = 1$ (squares). 
The dashed lines denote stress profiles
from DNS of pipe flow at $\Rey_\tau = 180$ (a,b,d) and $\Rey_\tau=360$ (c)~\citep{khoury_13}. 
           }
  \label{fig:tau11_trett}
 \end{center}
\end{figure}
 
The mean streamwise velocity field in compressible flows is generally characterized reverting to
the concept of compressibility transformations. \citet{morkovin_62}
first postulated that if density fluctuations are negligible
with respect to mean density variations, the direct effect of
compressibility on turbulence reduces to variations of the mean thermodynamic properties. 
This led to the well known van Driest transformation~\citep{vandriest_51},
which is quite accurate for adiabatic walls, whereas it is known to fail for isothermal 
walls~\citep{modesti_16}.
For the latter wall conditions, \citet{trettel_16}
have recently derived a compressibility transformation for channel flow
which relies on mean momentum balance and log-law universality.
\citet{pirozzoli_18} showed that in incompressible square duct flow
the streamwise velocity field is mainly influenced by the nearest wall, thus the wall law applies
with reasonable accuracy up to the corner bisector.
Based on this result and on the fact that the secondary motions
are not affected by compressibility, we then introduce the TL transformation 
for $y$ (the direction normal to the nearest wall) and $u$,
\begin{equation}
y_T(y, z)=\int_0^y f_T(\eta,z) \, {\mathrm d}\eta, \quad
u_T(y, z)=\int_0^y g_T(\eta,z) \, \frac{\partial \tilde{u}}{\partial \eta}(\eta, z) {\mathrm d}\eta,
\label{eq:TL}
\end{equation}
where the stretching functions are defined as
\begin{equation}
f_T(y,z)=\frac {\partial}{\partial y} \left( \frac y{R^{1/2} N} \right),\quad
g_T(y,z)=R N \frac {\partial}{\partial y} \left( \frac y{R^{1/2} N} \right),
\label{eq:kernels}
\end{equation} 
with $N(y,z)=\overline{\nu}/\overline{\nu}_w$, and $R(y,z)=\overline{\rho}/\overline{\rho}_w$.
It is important to note that the stretching functions $f_T$ and $g_T$ depend both on $y$ and $z$ (through the mean density
and mean viscosity), and likewise $y_T$ and $u_T$.

Figure~\ref{fig:uu_untr} shows the mean velocity profiles as a function
of the wall-normal distance up to the corner bisector (where mean velocity attains
a maximum), in local wall units. For reference purposes, the mean velocity profiles
from DNS of incompressible pipe flow are also reported~\citep{khoury_13}.
Excellent collapse of the velocity profiles at various $z$ is recovered,
also including the near-corner region. However, large differences are found 
among the different flow cases, especially at higher Mach number.
The TL transformed velocity profiles are shown
in figure~\ref{fig:uu_trett}, which 
now shows collapse of the velocity distributions 
both with respect to $z$ and across different flow cases.
This confirms on one hand that the TL transformation derived for plane channel
flow also holds with good approximation for square ducts.
On the other hand, the figure also suggests that close similarity with the 
velocity distributions in incompressible pipe flow is recovered for matching values
of an equivalent friction Reynolds number, which we define as
\begin{equation}
{\Rey_{\tau}}_T^* = y_T^*(h)/\delta_v^*. \label{eq:retaueq}
\end{equation}
where 
\begin{equation} 
y_T^*(y)= \frac 1h \int_{0}^h y_T(y,z)\mathrm{d}z, 
\end{equation}
is a stretched wall-normal coordinate averaged along the $z$ direction, 
which accounts in the mean for the variation of the local transformed scale $y_T$ with $z$.
As in the incompressible case therefore, duct flow shows close similarity with
pipe flow, which is a direct consequence of the fact that the intensity
of the secondary flows is rather small~\citep{pirozzoli_18}.

The streamwise turbulent stress ($\tau_{11}$) normalized by the local wall shear stress is shown
in figure~\ref{fig:tau11_trett}, and compared with incompressible pipe flow data. 
Along most of the wall, the behaviour is qualitatively similar to canonical
pipe flow, with a near-wall peak at $y^+ \approx 12$. The scatter among the various
$z$ sections appears to be generally much larger than for the mean velocity field, although
it seems to become confined to the corner vicinity at high enough $\Rey$.
The streamwise turbulent stress component exhibits a 
higher peak in the buffer layer at supersonic Mach number, which is not accurately captured by 
normalization with the local wall shear stress.
This effect was observed in previous studies of canonical compressible flows~\citep{coleman_95,ghosh_10,modesti_16},
but no convincing explanation has been provided to date.
Good collapse with incompressible pipe flow is by the way observed
for all cases in the outer part of the wall layer.
 
\subsection{Temperature field} \label{sec:temperature}
 
\begin{figure}
 \begin{center}
  \includegraphics[]{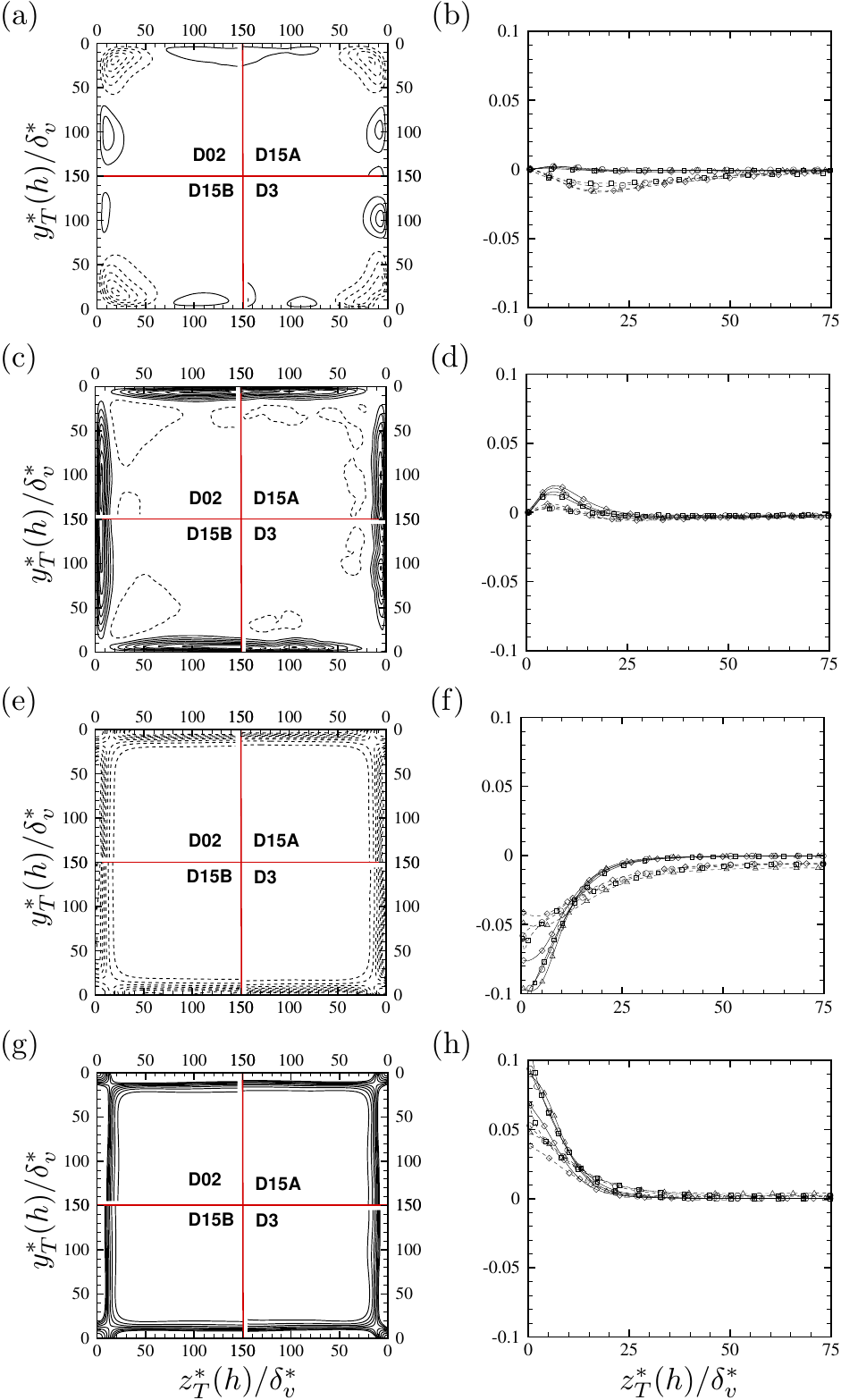}
  \vspace{1.em}
  \caption{Contours (panels a, c, e, g), and profiles (panels b, d, f, h) of mean enthalpy budget terms (equation~\eqref{eq:temp_duct}) in global stretched inner units. Contours are shown in the range $-0.025\le(.)/(\rho_w^* u_\tau^* T_\tau^*/\delta_v^*) \le 0.025$, in intervals of $3\cdot10^{-3}$, dashed lines denoting negative values. From top to bottom we show convection, turbulent transport, viscous diffusion, and viscous dissipation. The 1D profiles are reported at $y^*=25$ (dashed lines) and $y^*=75$ (solid lines), for all cases (D02, triangles; D15A, circles; D15B, diamonds; D3, squares).
           }
  \label{fig:temp_bud_in}
\end{center}
\end{figure}
 
\begin{figure}
 \begin{center}
  \includegraphics[]{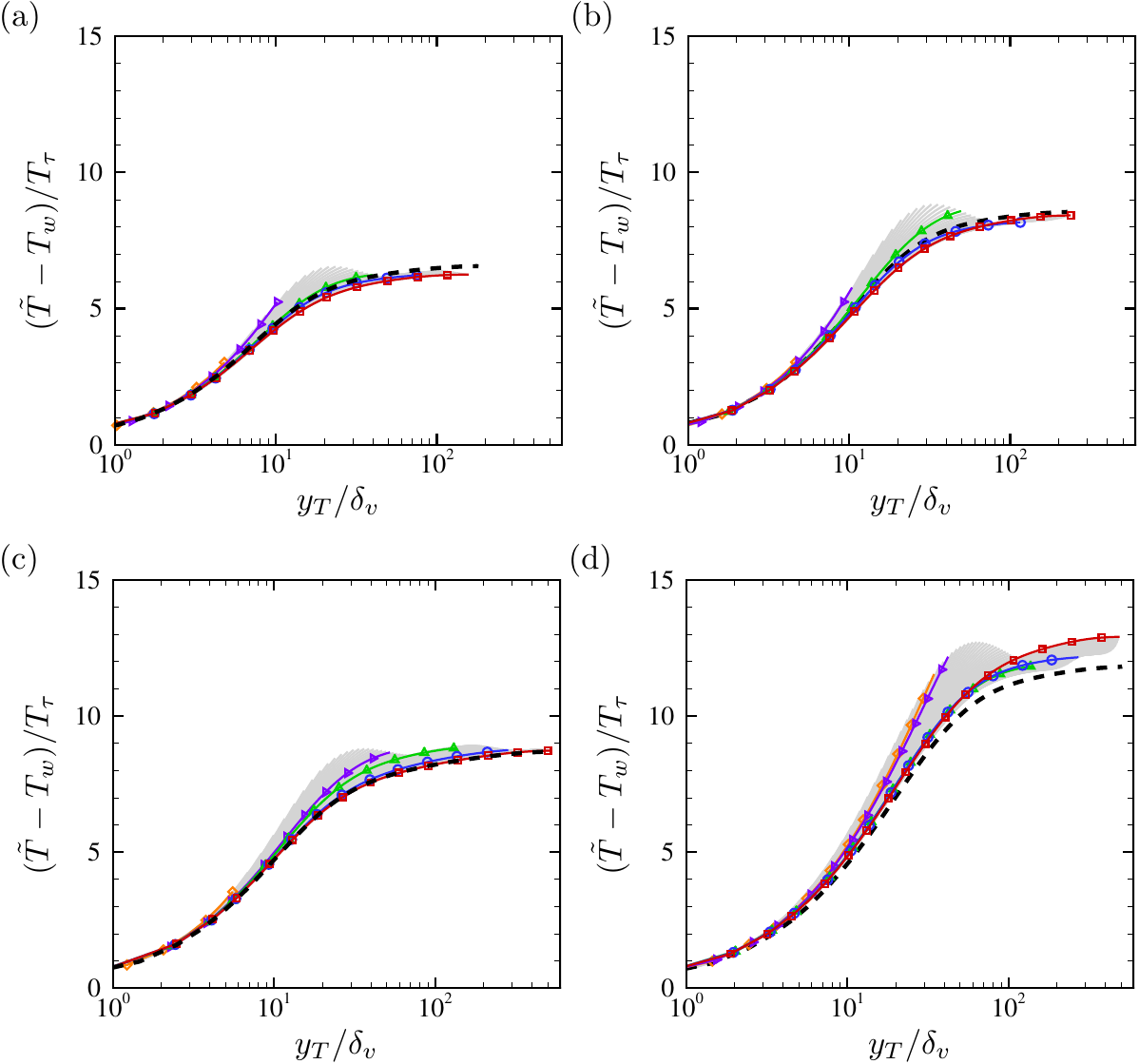}\\
  \vskip 1em
  \caption{Mean temperature profiles 
           along the $y$ direction (up to the corner
bisector), given in local wall units at all $z$,
for flow case D02 (${\Rey_{\tau}}_T^*=146$, panel a), D15A (${\Rey_{\tau}}_T^*=141$, panel b) 
D15B (${\Rey_{\tau}}_T^*=332$, panel c) and D3 (${\Rey_{\tau}}_T^*=145$, panel d).
Representative stations along the bottom wall
are highlighted, namely $z^* = 15$ (diamonds), $(z+h)/h = 0.1$ (right triangles), $(z+h)/h = 0.25$
(triangles), $(z+h)/h = 0.5$ (circles), $(z+h)/h = 1$ (squares). 
The dashed lines denote mean profiles
from DNS of pipe flow at $\Rey_\tau = 143$ (a,b,d) and $\Rey_\tau=334$ (c)~\citep{modestiphd_17}. 
The inset in panel (b) shows the mean streamwise velocity in the cross-stream plane with
symbols denoting representative $z$ stations.
           }
  \label{fig:tt}
 \end{center}
\end{figure}
 
\begin{figure}
 \begin{center}
%  (a)
  \includegraphics[width=15cm]{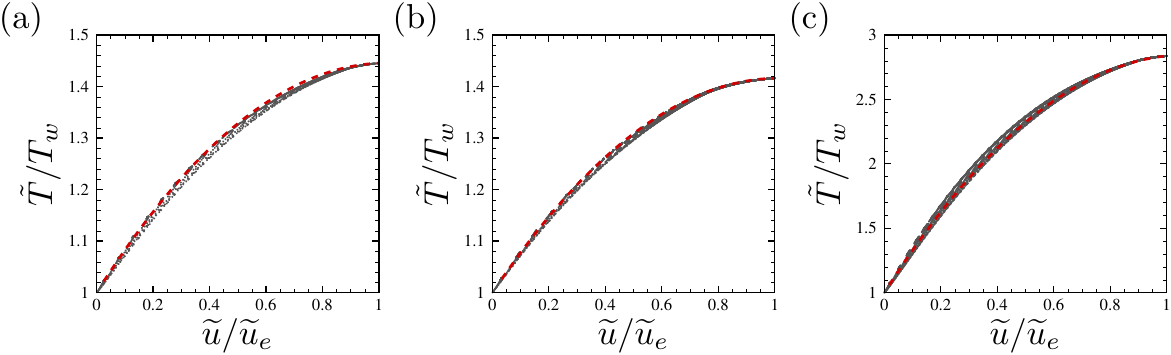}
  \vskip 1em
  \caption{Scatter plots of mean temperature versus mean velocity for all points in the duct cross section
           for flow cases D15A (a), D15B (b), D3 (c). The subscript $e$ refers to duct centerline values.
           The dashed line denotes the generalized temperature/velocity relation given in equation~\eqref{eq:zhang}.
           }
  \label{fig:tu_rel}
 \end{center}
\end{figure}

The temperature field is herein analyzed starting from the averaged enthalpy transport equation,
\begin{equation}
\begin{split}
\underbrace{\frac{\partial\overline{\rho}c_p\widetilde{v}\widetilde{T}}{\partial y} + 
\frac{\partial\overline{\rho}c_p\widetilde{w}\widetilde{T}}{\partial z}}_{C} &=  
\underbrace{\frac{\partial}{\partial y}\left(\overline{k}\frac{\partial\widetilde{T}}{\partial y}\right) + 
\frac{\partial}{\partial z}\left(\overline{k}\frac{\partial\widetilde{T}}{\partial z}\right)}_{D}+ \\ 
&\underbrace{ + \overline{u\frac{\partial p}{\partial x}} + \overline{v\frac{\partial p}{\partial y}} +  \overline{w\frac{\partial p}{\partial z}}}_{P}
+\underbrace{\overline{\sigma_{ij}\frac{\partial u_i}{\partial x_j}}}_{\Psi}
\underbrace{-\frac{\partial\overline{\rho}c_p\widetilde{v'' T''}}{\partial y}-\frac{\partial\overline{\rho}c_p\widetilde{w'' T''}}{\partial z}}_{T},
\end{split}
\label{eq:temp_duct}
\end{equation}
where the terms $C$, $D$, $P$, $\Psi$, and $T$ represent convection, viscous diffusion, 
pressure work, viscous dissipation and turbulent transport. The various contributions to the budget are reported in figure~\ref{fig:temp_bud_in},
in global stretched inner units, upon normalization of temperature with respect to the
global friction temperature, both in the cross-stream plane and at selected wall-normal sections.
The pressure work term is not reported, being negligible in all cases.
The figure shows that inner scaling yields good collapse across different Mach (cases D02, D15A, D3)
and Reynolds numbers (D15A, D15B).
Similar to what found from the mean momentum balance equation
in the incompressible case~\citep{pirozzoli_18},
we note that mean convection is mainly relevant at the duct corners, 
whereas it plays a minor role with respect to the other terms in the bulk region.
Viscous diffusion and dissipation contribute most to the budget, and they are
partially balanced by turbulent heat transport in the buffer layer.

Figure~\ref{fig:tt} shows the inner-scaled wall-normal temperature profiles up to the corner 
bisector, and it supports good universality in the spanwise direction, and good agreement between duct flow and supersonic pipe flow data at matching Mach and Reynolds number~\citep{modestiphd_17}.
Nevertheless, temperature does not exhibit any logarithmic layer nor universality with respect to Reynolds and Mach number,
unlike previously observed for the mean velocity field (see figure~\ref{fig:uu_trett}).
Corner effects also seem to be more significant than for the mean velocity, 
yielding earlier deviation from a common distribution when approaching the wall.
This behavior may be explained by noting that 
in the current case of isothermal wall the energy balance
is mainly controlled by aerodynamic heating, which is associated with viscous
dissipation, and which acts as a non-uniform spatial forcing. 
As shown in the forthcoming Section, in the case of a passive scalar with spatially
uniform forcing a logarithmic layer does in fact emerge as for the mean velocity.
Hence, deviations of the temperature distributions from a logarithmic behavior 
are the likely consequence of non-uniform heating, and regarding the 
temperature as a passive scalar may lead to incorrect conclusions, even at low Mach numbers.
 
Knowledge of the temperature distribution in compressible flow is necessary for the prediction
of friction~\citep{smits_06}. In particular, based on the distribution of $T$ one can derive 
the mean velocity distribution through reverse application of the
compressibility transformation introduced in Section~\ref{sec:transformations},
and temperature/velocity relationships are needed for the purpose.
The classical temperature/velocity relation by~\citet{walz_59} 
has proven its accuracy in the case of adiabatic walls~\citep{duan_10}, 
whereas it is found to fail in the case of isothermal walls~\citep{modesti_16}. 
Recently, \citet{zhang_14} derived the following generalized temperature/velocity relation,   
\begin{equation}
\frac{\widetilde{T}}{T_w} = 1 + \frac{T_{rg}-T_w}{T_w}\frac{\widetilde{u}}{\widetilde{u}_e} 
 + \frac{\widetilde{T}_e-T_{rg}}{T_w}\left(\frac{\widetilde{u}}{\widetilde{u}_e}\right)^2,
\label{eq:zhang}
\end{equation}
where $T_{rg}=\tilde{T}_e + r_g \tilde{u}_e^2/(2c_p)$ is a generalized recovery temperature,
$r_g = 2 c_p (T_w-\tilde{T}_e) / \tilde{u}_e^2 - 2 \Pran q_w / (\tilde{u}_e \tau_w)$ is a generalized recovery factor,
and $u_e$ and $T_e$ are the external values of velocity and temperature,
here interpreted as the duct centerline values.
Equation~\eqref{eq:zhang} explicitly takes into account
the wall heat flux $q_w$, and it reduces to Walz relation
in the case of adiabatic walls.
Figure~\ref{fig:tu_rel} shows scatter plots of temperature as a function of 
velocity for all points in the duct cross section.
Nearly perfect collapse of the supersonic DNS data with
the predictions of equation~\eqref{eq:zhang} is observed,
hence from integration of equation~\eqref{eq:kernels} one can 
reconstruct the full velocity field for assigned values of the bulk Reynolds and 
Mach numbers.

\subsection{Passive scalar field} \label{sec:passive}

\begin{figure}
 \begin{center}
  \includegraphics[]{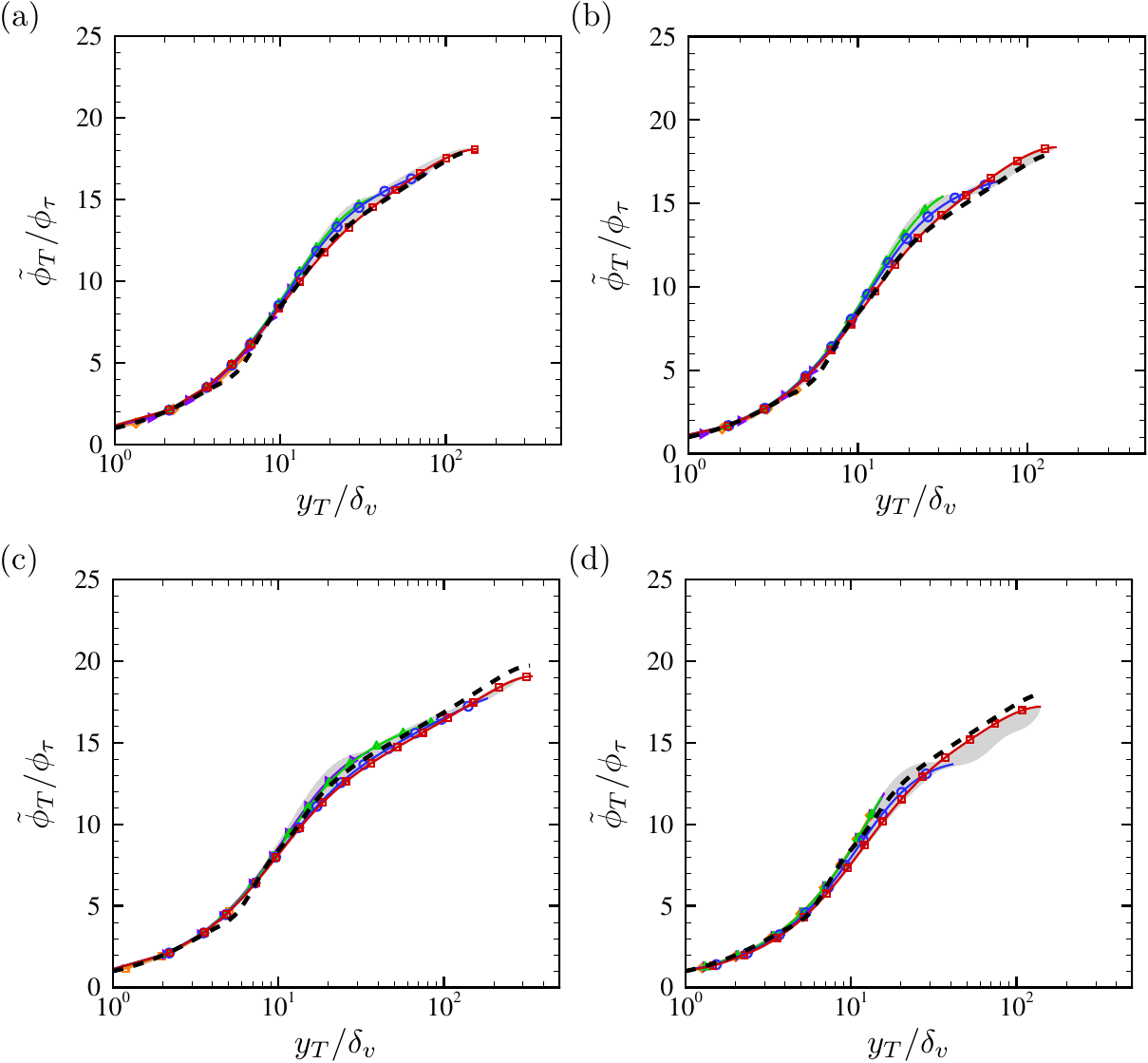}
  \vskip 1em
  \caption{Mean TL-transformed passive scalar profiles 
           along the $y$ direction (up to the corner
bisector), given in local wall units at all $z$,
for flow case D02 (${\Rey_{\tau}}_T^*=146$, panel a), D15A (${\Rey_{\tau}}_T^*=141$, panel b) 
D15B (${\Rey_{\tau}}_T^*=332$, panel c) and D3 (${\Rey_{\tau}}_T^*=145$, panel d).
$\phi_{\tau}$ is the friction value of the passive scalar, defined in equation~\eqref{eq:phitau}.
Representative stations along the bottom wall
are highlighted, namely $z^* = 15$ (diamonds), $(z+h)/h = 0.1$ (right triangles), $(z+h)/h = 0.25$
(triangles), $(z+h)/h = 0.5$ (circles), $(z+h)/h = 1$ (squares). 
The dashed lines denote fits of experimental incompressible pipe flow data by~\citet{kader_81}.
Different panels show flow cases D02 (a), D15A (b), D15B (c), D3 (d).}
  \label{fig:sca_prof}
 \end{center}
\end{figure}
\begin{figure}
 \begin{center}
%  (a)
  \includegraphics[width=15cm]{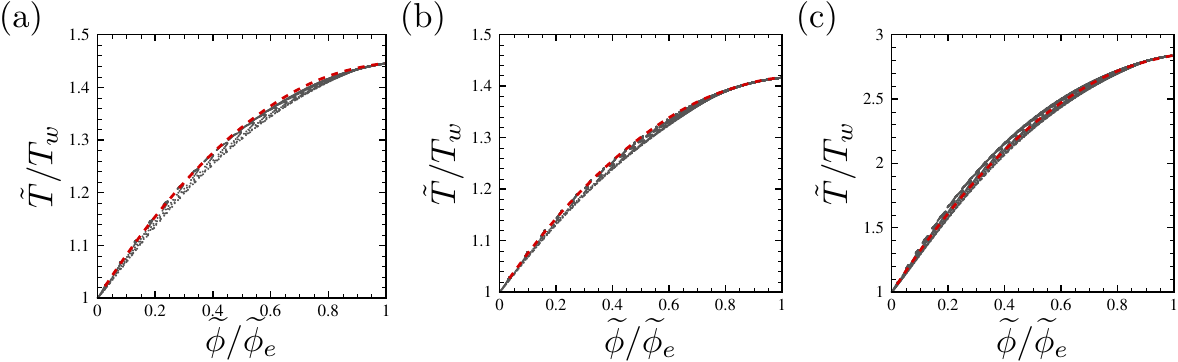}
  \vskip 1em
  \caption{Scatter plots of mean temperature versus mean passive scalar for all points in the duct cross section
           for flow cases D15A (a), D15B (b), D3 (c). The subscript $e$ refers to duct centerline values.
           The dashed line denotes the generalized temperature/passive scalar relation given in equation~\eqref{eq:zhang_sca}.
           }
  \label{fig:tsca_rel}
 \end{center}
\end{figure}

In this section we study the transport of a passive scalar field as from equation~\eqref{eq:scalar} at unit Schmidt number,
previously studied in channel flows in the incompressible case~\citep{pirozzoli_16}, and
for supersonic Mach number~\citep{foysi_05}.
Exploiting the similarity between the governing equation of a passive scalar (equation~\eqref{eq:scalar}) and the
streamwise momentum equation (equation~\eqref{eq:momentum}), we introduce a transformation for the mean scalar field
which mimics the TL transformation (equation~\eqref{eq:TL}), as follows
\begin{equation}
\phi_T(y, z)=\int_0^y g_T(\eta,z) \, \frac{\partial \tilde{\phi}}{\partial \eta}(\eta, z) \, {\mathrm d}\eta,
\label{eq:phitransf}
\end{equation}
with $g_T$ given in equation~\eqref{eq:kernels}. 
Figure~\ref{fig:sca_prof} shows the inner-scaled transformed passive scalar profiles, 
compared with the correlations developed by~\citet{kader_81} for incompressible pipe flow,
which include an inertial layer with logarithmic dependence on the wall distance.
The figure does in fact show accurate collapse of the $\phi_T$ distributions 
with respect to $z$ across the different flow cases and
with the incompressible fittings.
This findings supports our conjecture that the TL transformation can be extended 
to predict the behavior of passive scalars in compressible flow.
Again based on the close similarity between $\phi$ and $u$, 
we consider a generalization of equation~\eqref{eq:zhang} to relate temperature and passive scalars, namely
\begin{equation}
\frac{\tilde{T}}{T_w} = 1 + \frac{T_{rg}-T_w}{T_w}\frac{\widetilde{\phi}}{\widetilde{\phi}_e} 
 + \frac{\widetilde{T}_e-T_{rg}}{T_w}\left(\frac{\widetilde{\phi}}{\widetilde{\phi}_e}\right)^2 ,
\label{eq:zhang_sca}
\end{equation}
where $\widetilde{\phi}_e$ denotes the mean value of the passive scalar at the duct centerline.
Figure~\ref{fig:tsca_rel} shows the scatter plots of temperature as a function of 
the mean passive scalar for all points in the duct cross section.
As for the velocity field, near perfect collapse of the supersonic DNS data with
the predictions of equation~\ref{eq:zhang_sca} is observed,
which leads us to conclude that from integration of equation~\eqref{eq:phitransf} one can 
reconstruct the full passive scalar field for assigned values of the bulk Reynolds and 
Mach numbers.
 
\section{Conclusions}

We have carried out DNS of compressible square duct flow at various
Mach and Reynolds numbers, with the aim of clarifying the behavior of the
mean velocity, temperature and passive scalar fields.
All the flow cases exhibit the same typical secondary flow structure including eight counter-rotating eddies, 
which act to supply momentum to the duct corners.
The cross-stream velocity peaks are found to scale reasonably well with the bulk flow velocity, 
regardless of Mach and Reynolds number, with maximum value of about $2\%u_b$, which
is also consistent with the incompressible case~\citep{pirozzoli_18}.
The analysis of the mean streamwise velocity field leads to conclude
that the TL compressibility transformation derived for planar channel flow~\citep{trettel_16}
can be extended to square ducts, upon local application in the 
direction normal to the nearest wall.
The DNS data further suggest close similarity of the transformed velocity distributions with those
in incompressible pipe flow at matching values of an equivalent friction Reynolds number,
defined in equation~\eqref{eq:retaueq}.

Regarding the temperature field, we find that the various contributions to the enthalpy budget 
scale well when expressed in inner units, irrespective of the Mach and Reynolds numbers.
Similar to the incompressible case, we find that
mean convection is mainly relevant at the duct corners, 
whereas it plays a minor role with respect to the other terms in the bulk region.
Viscous diffusion and dissipation contribute most to the budget, and they are
partially balanced by turbulent heat transport in the buffer layer.
The energy balance is mainly controlled by aerodynamic heating, 
which acts as a non-uniform spatial forcing, 
and consequently the mean temperature does not exhibit any logarithmic layer nor universality with respect to Reynolds and Mach number.

Exploiting the similarity between the governing equation of a passive scalar and the
streamwise momentum equation, we show that the TL transformation can also be extended to 
predict the behavior of passive scalars in compressible flow, for which a logarithmic layer is found.
We further show that a generalized form of Waltz' equation can be used to relate the
mean velocity and passive scalar fields with the mean temperature field.
These results together point to a relatively simple representation of the mean flow properties 
in compressible duct flow, which can be exploited to obtain predictive formulas
for friction and heat transfer for assigned values of the bulk Reynolds and Mach numbers.

{\bf Acknowledgements}\\
We acknowledge that most of the results reported in this paper have been achieved using the PRACE Research Infrastructure resource MARCONI based at CINECA, Casalecchio di Reno, Italy.

\bibliographystyle{plainnat}
\bibliography{references}

\begin{thebibliography}{29}
\providecommand{\natexlab}[1]{#1}
\providecommand{\url}[1]{\texttt{#1}}
\expandafter\ifx\csname urlstyle\endcsname\relax
  \providecommand{\doi}[1]{doi: #1}\else
  \providecommand{\doi}{doi: \begingroup \urlstyle{rm}\Url}\fi

\bibitem[Bradshaw(1987)]{bradshaw_87}
P.~Bradshaw.
\newblock Turbulent secondary flows.
\newblock \emph{Annu. Rev. Fluid Mech.}, 19\penalty0 (1):\penalty0 53--74,
  1987.

\bibitem[Coleman et~al.(1995)Coleman, Kim, and Moser]{coleman_95}
G.N. Coleman, J.~Kim, and R.D. Moser.
\newblock A numerical study of turbulent supersonic isothermal-wall channel
  flow.
\newblock \emph{J.\ Fluid\ Mech.}, 305:\penalty0 159--183, 1995.

\bibitem[Davis et~al.(1986)Davis, Gessner, and Kerlick]{davis_86}
D.O. Davis, F.B. Gessner, and G.D. Kerlick.
\newblock Experimental and numerical investigation of supersonic turbulent flow
  through a square duct.
\newblock \emph{AIAA J.}, 24\penalty0 (9):\penalty0 1508--1515, 1986.

\bibitem[Duan et~al.(2010)Duan, Beekman, and Martin]{duan_10}
L.~Duan, I.~Beekman, and M.P. Martin.
\newblock Direct numerical simulation of hypersonic turbulent boundary layers.
  {P}art 2. {E}ffect of wall temperature.
\newblock \emph{J.\ Fluid\ Mech.}, 655:\penalty0 419--445, 2010.

\bibitem[{El Khoury} et~al.(2013){El Khoury}, Schlatter, Noorani, Fischer,
  Brethouwer, and Johansson]{khoury_13}
J.K. {El Khoury}, P.~Schlatter, A.~Noorani, P.F. Fischer, G.~Brethouwer, and
  A.V. Johansson.
\newblock Direct numerical simulation of turbulent pipe flow at moderately high
  {R}eynolds numbers.
\newblock \emph{Flow Turbul. Combust.}, 91\penalty0 (3):\penalty0 475--495,
  2013.

\bibitem[Foysi and Friedrich(2005)]{foysi_05}
H.~Foysi and R.~Friedrich.
\newblock {DNS} of {P}assive {S}calar {T}ransport in {T}urbulent supersonic
  channel flow.
\newblock In \emph{High Performance Computing in Science and Engineering,
  Munich 2004}, pages 107--117. Springer, 2005.

\bibitem[Gavrilakis(1992)]{gavrilakis_92}
S.~Gavrilakis.
\newblock Numerical simulation of low-{R}eynolds-number turbulent flow through
  a straight square duct.
\newblock \emph{J. Fluid Mech.}, 244:\penalty0 101--129, 1992.

\bibitem[Ghosh et~al.(2008)Ghosh, Sesterhenn, and Friedrich]{ghosh_08}
S.~Ghosh, J.~Sesterhenn, and R.~Friedrich.
\newblock Large-eddy simulation of supersonic turbulent flow in axisymmetric
  nozzles and diffusers.
\newblock \emph{Int. J. Heat Fluid Flow}, 29\penalty0 (3):\penalty0 579--590,
  2008.

\bibitem[Ghosh et~al.(2010)Ghosh, Foysi, and Friedrich]{ghosh_10}
S.~Ghosh, H.~Foysi, and R.~Friedrich.
\newblock Compressible turbulent channel and pipe flow: similarities and
  differences.
\newblock \emph{J. Fluid Mech.}, 648:\penalty0 155--181, 2010.

\bibitem[Kader(1981)]{kader_81}
B.A. Kader.
\newblock Temperature and concentration profiles in fully turbulent boundary
  layers.
\newblock \emph{Int. J. Heat Mass Transf.}, 24\penalty0 (9):\penalty0
  1541--1544, 1981.

\bibitem[Klein et~al.(2003)Klein, Sadiki, and Janicka]{klein_03}
M.~Klein, A.~Sadiki, and J.~Janicka.
\newblock A digital filter based generation of inflow data for spatially
  developing direct numerical or large eddy simulations.
\newblock \emph{J. Comput. Phys.}, 186\penalty0 (2):\penalty0 652--665, 2003.

\bibitem[Mani et~al.(2013)Mani, Babcock, Winkler, and Spalart]{mani_13}
M.~Mani, D.A. Babcock, C.M. Winkler, and P.R. Spalart.
\newblock Predictions of a supersonic turbulent flow in a square duct.
\newblock \emph{AIAA Paper}, 860:\penalty0 2013, 2013.

\bibitem[Modesti(2017)]{modestiphd_17}
D.~Modesti.
\newblock \emph{Direct numerical simulation of internal compressible flows at
  high Reynolds number: numerical and physical insight}.
\newblock PhD thesis, La Sapienza Unversit\`a di {R}oma, 2017.
\newblock Published online: http://hdl.handle.net/11573/939526.

\bibitem[Modesti and Pirozzoli(2016)]{modesti_16}
D.~Modesti and S.~Pirozzoli.
\newblock Reynolds and {M}ach number effects in compressible turbulent channel
  flow.
\newblock \emph{Int. J. Heat Fluid Flow}, 59:\penalty0 33--49, 2016.

\bibitem[Modesti and Pirozzoli(2018)]{modesti_18}
D.~Modesti and S.~Pirozzoli.
\newblock An efficient semi-implicit solver for direct numerical simulation of
  compressible flows at all speeds.
\newblock \emph{J. Sci. Comput.}, 75\penalty0 (1):\penalty0 308--331, 2018.

\bibitem[Morajkar and Gamba(2016)]{morajkar_16}
R.R. Morajkar and M.~Gamba.
\newblock Turbulence characteristics of supersonic corner flows in a low aspect
  ratio rectangular channel.
\newblock In \emph{54th AIAA Aerospace Sciences Meeting}, page 1590, 2016.

\bibitem[Morkovin(1962)]{morkovin_62}
M.V. Morkovin.
\newblock Effects of compressibility on turbulent flows.
\newblock In \emph{M{\'e}canique de la Turbulence}, pages 367--380. A. {F}avre,
  1962.

\bibitem[Nikuradse(1926)]{nikuradse_26}
J.~Nikuradse.
\newblock \emph{Untersuchung {\"u}ber die Geschwindigkeitsverteilung in
  turbulenten Str{\"o}mungen}.
\newblock Vdi-verlag, 1926.

\bibitem[Pirozzoli(2010)]{pirozzoli_10}
S.~Pirozzoli.
\newblock Generalized conservative approximations of split convective
  derivative operators.
\newblock \emph{J. Comput Phys.}, 229\penalty0 (19):\penalty0 7180--7190, 2010.

\bibitem[Pirozzoli et~al.(2016)Pirozzoli, Bernardini, and
  Orlandi]{pirozzoli_16}
S.~Pirozzoli, M.~Bernardini, and P.~Orlandi.
\newblock Passive scalars in turbulent channel flow at high {R}eynolds number.
\newblock \emph{J.~Fluid~Mech.}, 788:\penalty0 614--639, 2016.

\bibitem[Pirozzoli et~al.(2018)Pirozzoli, Modesti, Orlandi, and
  Grasso]{pirozzoli_18}
S.~Pirozzoli, D.~Modesti, P.~Orlandi, and F.~Grasso.
\newblock Turbulence and secondary motions in square duct flow.
\newblock \emph{J. Fluid Mech.}, 840:\penalty0 631--655, 2018.

\bibitem[Prandtl(1927)]{prandtl_27}
L.~Prandtl.
\newblock {\"U}ber den reibungswiderstand str{\"o}mender luft.
\newblock \emph{Reports of the Aerod. Versuchsanst, G{\"o}ttingen, Germany, 3rd
  Series}, 1927.

\bibitem[Smits and Dussauge(2006)]{smits_06}
A.J. Smits and J.-P. Dussauge.
\newblock \emph{Turbulent Shear Layers in Supersonic Flow}.
\newblock 2nd edn., American Institute of Physics, New York, 2006.

\bibitem[Trettel and Larsson(2016)]{trettel_16}
A.~Trettel and J.~Larsson.
\newblock Mean velocity scaling for compressible wall turbulence with heat
  transfer.
\newblock \emph{Phys. Fluids (1994-present)}, 28\penalty0 (2):\penalty0 026102,
  2016.

\bibitem[van Driest(1951)]{vandriest_51}
E.R. van Driest.
\newblock Turbulent boundary layer in compressible fluids.
\newblock \emph{J.\ Aero.\ Sci.}, 18:\penalty0 145--160, 1951.

\bibitem[Vane and Lele(2015)]{vane_15}
Z.P. Vane and S.K. Lele.
\newblock Prediction of {T}urbulent {S}econdary {F}lows in {D}ucts {U}sing
  {E}quilibrium {W}all-{M}odeled {LES}.
\newblock In \emph{53rd AIAA Aerospace Sciences Meeting}, page 1274, 2015.

\bibitem[V{\'a}zquez and M{\'e}tais(2002)]{vazquez_02}
M.S. V{\'a}zquez and O.~M{\'e}tais.
\newblock Large-eddy simulation of the turbulent flow through a heated square
  duct.
\newblock \emph{J. Fluid Mech.}, 453:\penalty0 201--238, 2002.

\bibitem[Walz(1959)]{walz_59}
A.~Walz.
\newblock Compressible turbulent boundary layers with heat transfer and
  pressure gradient in flow direction.
\newblock \emph{J.\ Res.\ Natl.\ Bur.\ Stand.}, 63, 1959.

\bibitem[Zhang et~al.(2014)Zhang, Bi, Hussain, and She]{zhang_14}
Y.S. Zhang, W.T. Bi, F.~Hussain, and Z.S. She.
\newblock A generalized {R}eynolds analogy for compressible wall-bounded
  turbulent flows.
\newblock \emph{J. Fluid Mech.}, 739:\penalty0 392--420, 2014.

\end{thebibliography}
 
\end{document}